# Feasible Rotator Using Zero Index Metamaterials and Perfect Electric Conductors

M. M. Sadeghi [1, *], H. Nadgaran [2]

[1] *Department of Physics, Jahrom University, Jahrom 74137-66171, Iran*

[2] *Department of Physics, Shiraz University, Shiraz 71454, Iran*

\* E-mail: sadeghi@jahromu.ac.ir

**Abstract:** We report the design of a new electromagnetic device with a new mapping function to have simultaneous electromagnetic concentration and rotation using a singular radial mapping. We implement such a device only by using alternating structure of zero index metamaterials and perfect electric conductors. Numerical simulations are performed to verify its functionality.

1. **Introduction**

Transformation optics was first proposed by Leonhardt and Pendry [1, 2]. In this theory they used coordinate transformation to convert constitute parameters in some specific region to yield invisibility cloak. Soon afterward a lot of attention has been paid to this theory [3-7] and by using this theory a lot of various and fascinating optical devices was designed and implemented. In this respect theoretical and experimental works such as optical black holes [8, 9], super lenses [10, 11], anti-cloak [12, 13] and [14, 15] were reported.

Among various optical devices, EM field rotators is an important device promising significant potential application in illusion devices. In 2007 Chen et.al [16] proposed the first rotator designed by transformation optics. In 2009, H. Y. Chen et al. [17] could get the first realization of field rotators at microwave frequencies. One can find more works on field rotators [18–21].

In 2015 Sadeghi et.al [22] employed a novel structure to implement Fabry-Perot (FP)-based devices where a three-fold concerns should be noticed in this work. First the structure has been working for those frequencies that was supported by a FP cavity with lengths as a multiple integer of the wavelengths that the paper were working at. In other words one cannot use the device for a continuous range of frequencies. Second the proposed structure for rotators was not capable of bearing a simple affordable construction and manufacture for curved plates surrounding the core medium. Third, based on extreme constitute parameters in r direction resulting from singular transformation function, one cannot achieve promising results (see fig.5b of present paper).

In this paper we propose a rotator that not only works for desired frequencies without limitation, but still uses ZIMs and PECs similar to those in our previous work on perfect concentrators [23]. In other words we tried to design rotators with improvements in the transformation function of the core region to avoid great impedance mismatch in core boundary.

Here a review of the concentrators and rotator concepts will be given. Then the reconstructed mapping transformation function will be presented and based on this transformation function the related constitute parameters and the realization of a device similar to those reported for perfect concentrators will be discussed [23].

## 2. Concentrators

We start from concentrator transformation function with radial transformation similar to those in [5] that schematically was shown in fig.1.

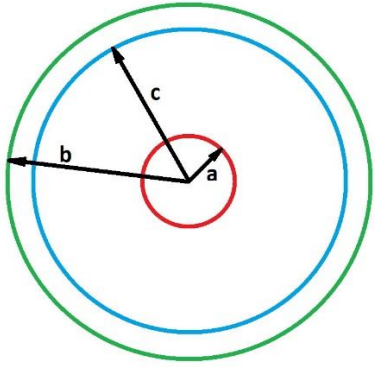

Fig.1: The schematic plot of proposed mapping

For $0 \leq r' \leq c$, $r = f(r') = \left(\frac{a}{c}\right) r'$

For $c \leq r' \leq b$, $r = f(r') = \left(\frac{b-a}{b-c}\right) r' - \left(\frac{c-a}{b-c}\right) b$

And for $r' \geq b$, $r = r'$

This mapping function compress region $r' \in [0, c]$ into $r \in [0, a]$ and expand region $r' \in [c, b]$ to $r \in [a, b]$. So according to transformation optics theory the permittivity and permeability can be derived as:

For $0 \leq r \leq a$,

$$\varepsilon = \mu = \begin{pmatrix} 1 & 0 & 0 \\ 0 & 1 & 0 \\ 0 & 0 & \left(\frac{c}{a}\right)^2 \end{pmatrix} \quad (1)$$

And for $a \leq r \leq b$,

$$\varepsilon = \mu = \begin{pmatrix} \eta_r & 0 & 0 \\ 0 & \eta_r^{-1} & 0 \\ 0 & 0 & \left(\frac{f}{h}\right)^2 \eta_r \end{pmatrix} \quad (2)$$

where $\eta_r = \frac{e}{f} \frac{b}{r} + 1$, $e = c - b$, $f = b - c$, and $h = c - a$. Now if we increase $c$ so that $c = b$, the transformation will become singular. In fig.2a we illustrate full-wave simulations to visualize properties of this

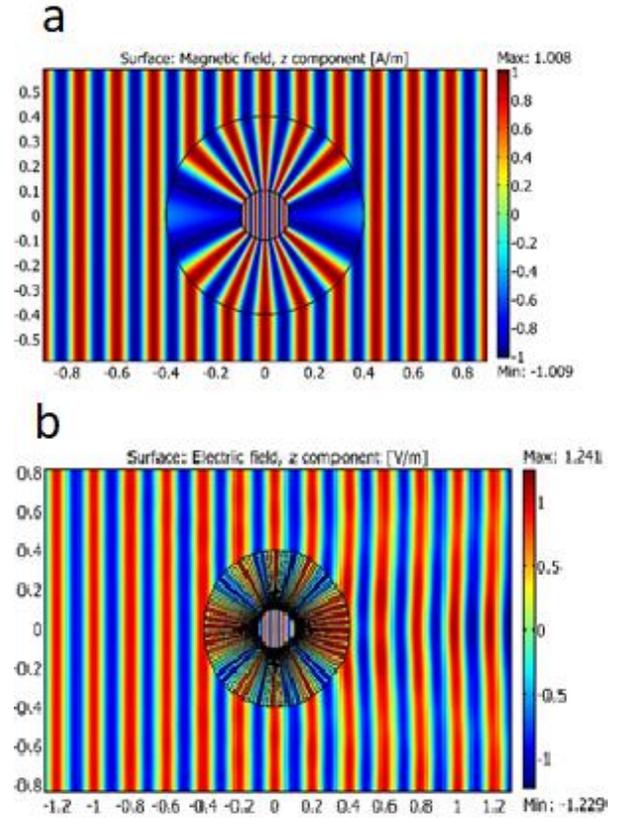

Fig2: TM plane wave with frequency of 2GHz incident on the device with the inner radius $a$=0.1 m and the outer radius $b$=0.4 m. a. concentrator with effective medium parameters b. concentrator with proposed structure.

media for a TM plane wave with frequency of 2GHz, the inner radius $a$=0.1 m and the outer radius $b$=0.4 m.

For implementation such singular invisible device similar to paper [23] we here consider the periodic structure composing of two kind of isotropic metamaterials. These two kind of isotropic metamaterials are excellent candidate for singular radial mapping, so we arranged them in alternating radial slices as we can see in fig.1b.

### 3. Rotators

Now we focus on field rotator proposed by ref. [16] with the following mapping function.

For $r < a$, $r' = r$, $z' = z$, $\theta' = \theta + \theta_0$ (3)

For $r > b$, $r' = r$, $z' = z$, and $\theta' = \theta$ (4)

and for $a < r < b$,

$r' = r$, $z' = z$, $\theta' = \theta + \theta_0 \frac{f(b)-f(r)}{f(b)-f(a)}$ (5)

Which, schematically using fig. 1, for the inner cylinder, rotates the field with an angle $\theta_0$. The electric permittivity and magnetic permeability tensors in this case can be written as:

$\varepsilon_{xx} = 1 + 2t \cos\theta' \sin\theta' + t^2 \sin^2\theta'$
$= \varepsilon_u \cos^2(\theta' + \tau/2)\varepsilon_v \sin^2(\theta' + \tau/2)$

$\varepsilon_{xy} = \varepsilon_{yx} = -t^2 \cos\theta' \sin\theta' - t(\cos^2\theta' - \sin^2\theta')$
$= (\varepsilon_u - \varepsilon_v)\sin(\theta' + \tau/2)\cos(\theta' + \tau/2)$

$\varepsilon_{xx} = 1 - 2t \cos\theta' \sin\theta' + t^2 \cos^2\theta' = \varepsilon_u \sin^2\left(\theta' + \frac{\tau}{2}\right) + \varepsilon_v \cos^2\left(\theta' + \frac{\tau}{2}\right)$

Where,

$\varepsilon_u = 1 + \left(\frac{1}{2}\right)t^2 - \left(\frac{1}{2}\right)t\sqrt{t^2 + 4}$,

$\varepsilon_v = 1 + \left(\frac{1}{2}\right)t^2 + \left(\frac{1}{2}\right)t\sqrt{t^2 + 4}$

Here we let $t = \frac{\theta_0 r f'(r)}{f(b)-f(a)} = \frac{\theta_0}{\ln\left(\frac{b}{a}\right)}$ and $f(r) = r$ similar to those in ref. [16].

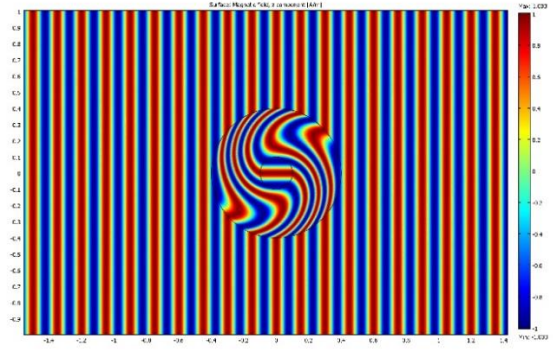

Fig3. TM plane wave with frequency of 2GHz, the inner radius $a$=0.1 m and the outer radius $b$=0.4 m

By using incident TM wave to visualize properties of this media with frequency of 2GHz, the inner radius $a$=0.1 m, the outer radius $b$=0.4 m and $\theta_0 = 15$, we end up with fig. 3.

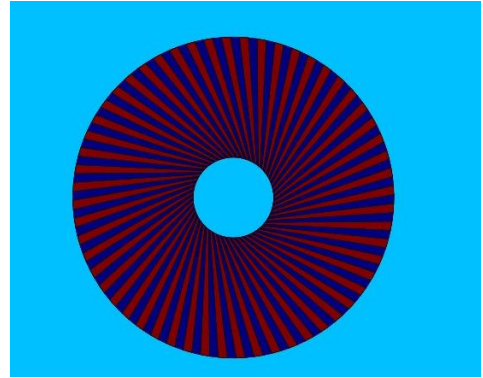

Fig. 4: proposed structure for rotator

For implementation of such a rotator if we use the previous ZIMs and PECs, we cannot get desirable results (see fig.5a). In fact the effective constitutive of such a periodic radial structure with extreme constitute parameters in r direction, squeezes space from a surrounding volume into a core volume while one cannot conclude this squeeze and concentration based on transformation function in core region (see equation(3)). In fact if we use such a structure it means we have non-continuous transformation function in core region and hence we have great mismatch impedance in core boundary. In fig. 5a and 5b we can see the huge scattering due to impedance mismatch in core boundary for effective medium and proposed structure. To cope with this mismatch in core boundary we introduce a new transformation function to overcome this impedance mismatch.

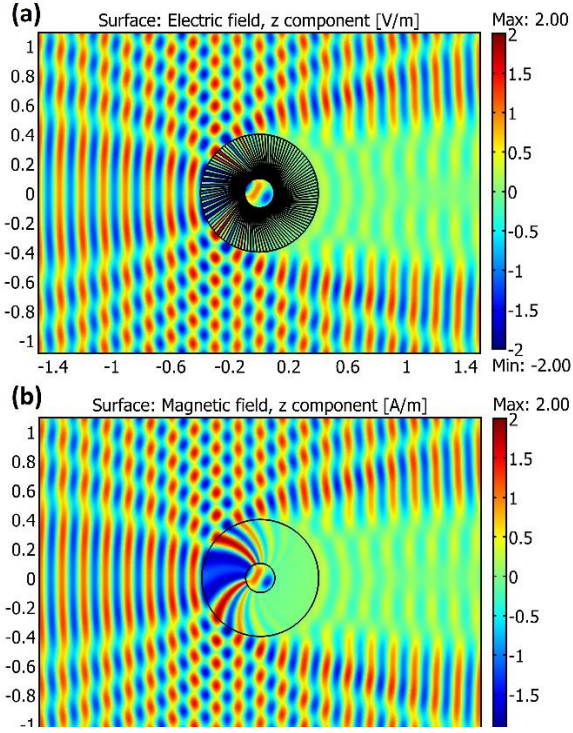

Fig 5a: the TM field distribution for configuration with 112 slices with angle of 15 degree tilted slices. b. the TM field distribution for effective medium

## 4. Perfect Rotators

Here we report a new mapping transformation function to satisfy continuity of radial transformation function in core boundary and rotation in core region.

For $r > b$, $r' = r$, $z' = z$, and $\theta' = \theta$

For $r < a$, $r' = f(r) = \left(\frac{a}{c}\right)r$, $z' = z$, $\theta' = \theta + \theta_0$

and for $a < r < b$, $r = \frac{b-a}{b-c}r' - \frac{R_2-a}{b-c}b$

$\theta = \theta' + \theta_0 \frac{f(b)-f(r')}{f(b)-f(a)}$, $z = z'$

This mapping function compressed region $r' \in [0,c]$ into $r \in [0,a]$ and expand region $r' \in [c,b]$ to $r \in [a,b]$ and for the inner cylinder rotates the field an angle $\theta_0$. Base on transformation optics theory this transformation function result in permittivity and permeability tensors as:

$\varepsilon_{rr} = (r+\beta)/r$

$\varepsilon_{r\theta} = -k\frac{(r+\beta)}{(\alpha r)}$

$\varepsilon_{\theta\theta} = \frac{(r+\beta)}{(\alpha^2 r)}k^2 + \frac{r}{(r+\beta)}$

$\varepsilon_{zz} = \frac{(r+\beta)}{(\alpha^2 r)}$

Where $k = \frac{r}{b-a}\theta_0$, $\beta = \frac{c-a}{b-c}b$ and $\alpha = \frac{b-a}{b-c}$.

In Cartesian coordinates these tensors components can be written as

For $0 \leq r \leq a$,

$$\varepsilon = \mu = \begin{pmatrix} 1 & 0 & 0 \\ 0 & 1 & 0 \\ 0 & 0 & \left(\frac{c}{a}\right)^2 \end{pmatrix}$$

And for $a \leq r \leq b$,

$\varepsilon_{xx} = \varepsilon_{rr}\cos^2\theta - 2\varepsilon_{r\theta}\sin\theta\cos\theta + \varepsilon_{\theta\theta}\sin^2\theta$

$\varepsilon_{xy} = \varepsilon_{rr}\sin\theta\cos\theta + \varepsilon_{r\theta}(\cos^2\theta - \sin^2\theta) - \varepsilon_{\theta\theta}\sin\theta\cos\theta$

$\varepsilon_{yy} = \varepsilon_{rr}\sin^2\theta - 2\varepsilon_{r\theta}\sin\theta\cos\theta + \varepsilon_{\theta\theta}\cos^2\theta$

$$\varepsilon_{zz} = \frac{(r+\beta)}{(\alpha^2 r)}$$

We illustrated magnetic field pattern for z component in fig. 6 to visualize properties of this kind of transformation media for an incident TM plane wave with frequency of 2GHz, the inner radius is $a$=0.1 m and outer radius $b$=0.4 m while rotation angle is $\theta_0 = 15$. As we can see in this figure, in this device the incident field has been concentrated and rotated simultaneously in core medium.

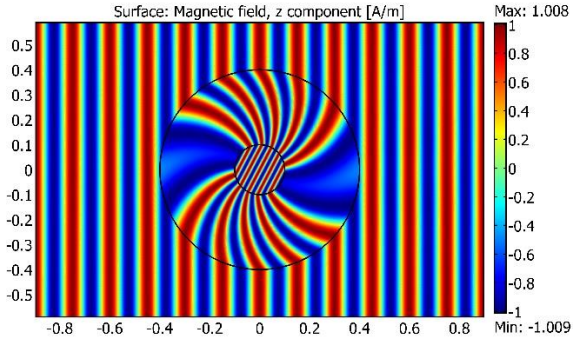

Fig6: TM magnetic field pattern with working frequency of 2GHz, the inner radius $a$=0.1 m and the outer radius $b$=0.4 m

Now we introduce a way to implement such singular invisible device. Similar to Ref. [23] we consider the periodic structure composing of two kind of isotropic metamaterials in order to concentrate and rotate fields simultaneously. These two kind of isotropic metamaterials are excellent candidate for singular radial mapping, so we arranged them in alternating tilted slices as we can see in fig7.

Now to prove and visualize properties of such a proposed structure full wave simulations in two dimension have been performed to verify its functionality and its illusionary effects for TM-z polarization incident wave. At first we plot the TM field distribution in fig. 5a and 5b for configuration with 112 slices and 256 slices with angle of 15 degree tilted slices. Moreover working frequency again is 2GHz, inner radius is a=0.1 m while the outer radius is b=0.4 m. in this simulation we set $\varepsilon = -100000$ for PECs and $\varepsilon = \mu = 0.001$ for ZIMs.

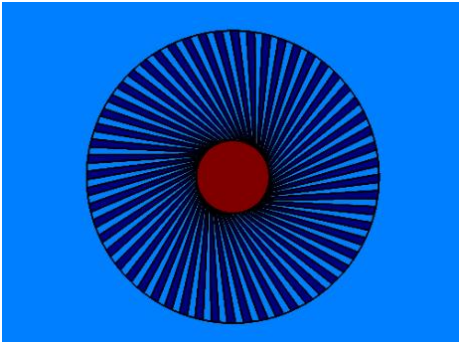

Fig 7: the proposed structure for $c \leq r' \leq b$ that comprised of two homogenous and isotropic metamaterials.

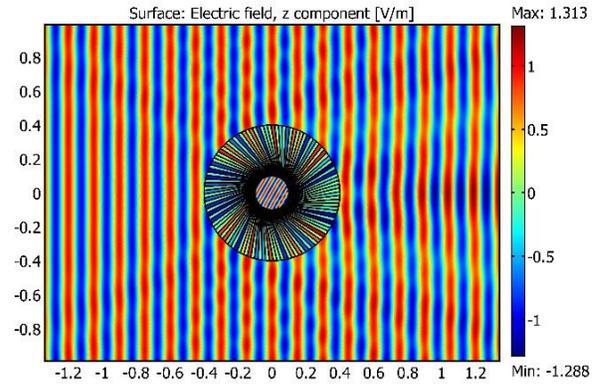

Fig 8: the TM field distribution for configuration with 112 slices and 256 slices with angle of 15 degree tilted slices.

From fig.8 we can see the result of concentration and rotation are the same as those in fig.6 that calculated by transformation optics media.

In summary, we first introduced a new mapping transformation function that result in a new perfect device and then by using transformation optics theory we calculated correspondence constitute parameters. Finally we proposed a structure using only two kind of isotropic and homogeneous metamaterials, ZIMs and PECs to have the same perfect effects.